# Improved nanopatterning for YBCO nanowires approaching the depairing current

Riccardo Arpaia, Shahid Nawaz, Floriana Lombardi, and Thilo Bauch

*Abstract*—An improved nanopatterning procedure has been developed to obtain YBa$_2$Cu$_3$O$_{7-x}$ (YBCO) nanowires with cross sections as small as 50x50 nm$^2$, protected by an Au capping layer. To probe the effective role of the Au protecting layer, we have measured the current-voltage characteristics and the resistive transition in temperature of the nanowires. Critical current densities up to 10$^8$ A/cm$^2$ have been achieved at T=4.2 K, approaching the theoretical depairing current limit. The resistance, measured as a function of temperature close to T$_c$, has been fitted with a thermal activated phase slip model, including the effect of the gold layer. The extracted values of the superconducting coherence length and of the London penetration depth give current densities consistent with the measured ones. These results cannot be achieved with same nanowires, without the Au capping layer.

## I. Introduction

THE improvement of nanopatterning techniques during recent years has enabled the study of fundamental aspects of superconductivity on the nanoscale [1], [2], such as thermal and quantum phase slip dynamics in nanowires giving rise to a finite resistance at temperatures below the superconducting transition temperature. Nanoscale superconductors also pave the way for new exciting developments towards quantum-limited sensors for single photons and spins [3], [4]. The realization of wires with highly homogeneous superconducting properties is essential to enable fundamental studies and reproducible devices. While the reliable fabrication of nanostructures is nowadays feasible for conventional superconductors [5], it still represents an extremely challenging task for cuprate High critical Temperature Superconductors (HTSs). The chemical instability of these materials, and the extreme sensitivity to defects and disorder due to the very short superconducting coherence length $\xi$ (of the order of 2 nm), make the establishment of a reliable nanofabrication procedure a challenging task.

Here we present an optimized nano-patterning process for YBa$_2$Cu$_3$O$_{7-x}$ (YBCO) nanowires with cross sections as small as $50 \times 50$ nm$^2$ implementing a protecting gold capping layer. We characterized the electrical transport properties both at temperatures close to the superconducting transition temperature, $T_c = 85$ K, and at $T = 4.2$ K. From the resistance as a function of temperature, $R(T)$, measured close to $T_c$ we fit the superconducting coherence length $\xi$ and the London penetration depth $\lambda_L$ using a thermal activated phase slip model [6], [7], including the effect of the gold capping. Differently to the analysis of $R(T)$ measurements performed on YBCO wires reported in literature [8], [9], [10], we provide additional corroboration of our fitting procedure by comparing the measured critical current densities of the same wires at $T = 4.2$ K to the theoretically expected Ginzburg-Landau critical current densities using the fitted values $\xi$ and $\lambda_L$.

## II. Fabrication

We have fabricated YBCO nanowires by depositing 50 nm thick film of YBCO, c-axis oriented, on a MgO (110) substrate by Pulsed Laser Deposition (PLD). The growth conditions were optimized to achieve the right compromise between critical temperature T$_c$ and smoothness of the surface. We obtained epitaxial thin films with a Tc of ~85 K and a 1 K wide transition, and an average surface roughness below 4 nm measured by Atomic Force Microscopy (AFM). The films were covered ex-situ, soon after the deposition, by a 200 nm thick gold film to prevent any oxygen loss and to avoid any possible contaminations. The Au pads were then defined by photolithography and Ar$^+$ ion etching, making sure to leave a 50 nm thick Au layer on the remaining chip areas, working as a cap layer for YBCO. To define the nanowires we have used a 100 nm thick amorphous carbon film as a hard mask, in combination with electron beam lithography at 100 kV and Ar$^+$ ion etching [9], [11]. This last step is quite crucial: the interaction of Ar$^+$ ions with YBCO can strongly affect the transport properties because of severe changes in film stoichiometry and disorder. This is even more remarkable while defining nanostructures: big differences can be observed in the transport properties, in particular the value of the critical current density and its temperature dependence, by tiny changes of the etching parameters. So, the achievement of reproducible results at the nanoscale lies ultimately in a good control of the etching process. In our case, we tuned the etching parameters to avoid overheating of the sample and we calibrated carefully the etching times in order to minimize the interaction of the Ar$^+$ ions with the nanowires. In particular, a key parameter is the accelerating voltage of the Ar$^+$ ion beam.

Manuscript received October 5, 2012. This work was has been partially supported by the Swedish Research Council (VR) and the Knut and Alice Wallenberg Foundation.

The authors are with the Quantum Device Physics Laboratory, Department of Microtechnology and Nanoscience, MC2, Chalmers University of Technology, Göteborg S-41296, Sweden (email: arpaia@chalmers.se).

R. Arpaia is also with CNR-SPIN, Dipartimento di Scienze Fisiche Università degli Studi di Napoli Federico II, Italy.

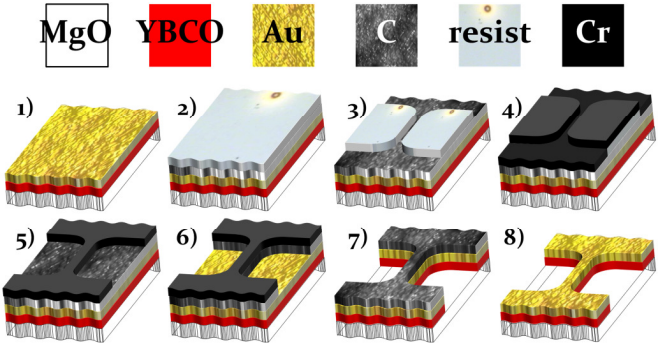

Fig. 1. Fabrication scheme: (1) Au/YBCO bilayer is grown on the substrate, (2) a carbon layer acting as a hard mask is evaporated and then a double layer resist is spun, (3) the resist is developed after the e-beam lithography, (4) a chromium layer is evaporated on top, (5) the mask is defined after Cr is lifted-off, (6) uncovered C is removed by oxygen plasma etching, (7) chip after ion-beam etching, (8) the nanowires are achieved after oxygen plasma removal of the residual C.

We have chosen to work close to the etching voltage threshold for YBCO: below this value, which is of the order of 300 V, the film is only made amorphous instead of being etched. In this paper we will discuss the transport properties of nanowires with a 50 nm thick gold cap layer protecting the nanostructures and we will compare them with those nanowires where the gold cap layer has been completely removed by a further $Ar^+$ etching procedure. Fig. 1 summarizes the main fabrication steps.

AFM and Scanning Electron Microscopy (SEM) have been used to determine for every nanowire the real dimensions of the nanostructures. We have achieved nanowires with widths down to 40 nm and lengths ranging between 200 nm and 3 um (Fig 2). In this paper we will focus on the transport properties of the shortest ones.

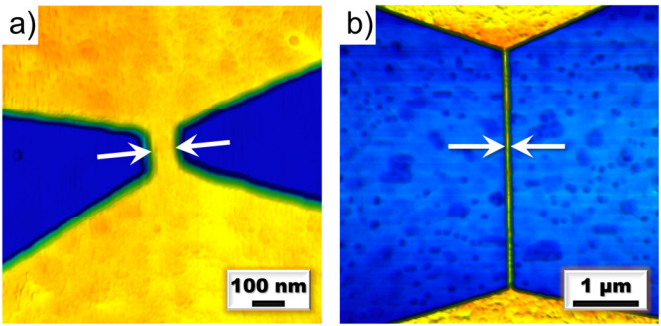

Fig. 2. AFM pictures of two nanowires, with the following dimensions: (a) width=65 nm, length=200 nm (b) width=40 nm, length=3 µm.

### III. DEVICE CHARACTERIZATION

The electrical transport properties of the nanobridges were measured in a $^3$He cryostat using a four probe configuration. Resistance versus temperature and current-voltage characteristic were carried out for nanowires with and without capping Au.

We measured in total 200 nanowires, belonging to ten different samples. In this paper, we will consider data obtained from two samples representative of the whole wire population.

Figure 3(a) shows the current-voltage characteristic (IVC)

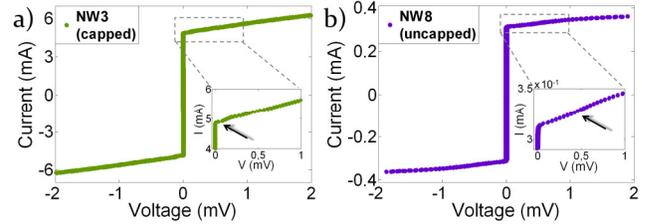

Fig. 3. Current-voltage characteristic (IVC) measured at T=4K for the Au capped wire NW3 (a) and for the uncapped wire NW8 (b). The two insets show a close-up of the resistive branches of the two IVCs close to the critical current Ic: the arrows indicate the steps explainable in terms of phenomena like phase slip centers or hot spots.

at 4.2 K for a nanowire 95 nm wide capped with Au. For comparison in Figure 3(b) it is reported the IVC of a nanowire with similar dimensions without the protecting Au. The critical current densities Jc for these two nanowires are $1 \times 10^8$ A/cm$^2$ and $2 \times 10^7$ A/cm$^2$, respectively. They represent the highest values measured for nanowires with and without capping Au. Most of the wires protected by gold show a Jc value similar to that of NW3 (Fig. 3(a)): such values of the critical current density are much larger than any values reported in literature [8] for YBCO and are close to the theoretical depairing current limit. As predicted by the Ginzburg-Landau (GL) theory, the expression for the depairing current density is [12]:

$$j_d^{GL}(T) = \Phi_0 / 3^{3/2} \pi \mu_0 \lambda^2(T) \xi(T) \qquad (1)$$

with $\Phi_0$ the flux quantum, $\mu_0$ the vacuum permeability, $\lambda(T)$ the London penetration depth and $\xi(T)$ the coherence length. Here it is important to note that this equation is strictly valid only at temperatures close to the critical one. However, calculations for the depairing critical current density in the clean and dirty limit at T=0 K predict a deviation from the GL limit by a maximum factor of 1.5 [13]. In the limit of low temperature, typical values for the London penetration depth and coherence length for (001) YBCO films are $\lambda_{ab}(0) = 150 \div 250$ nm [8] and $\xi_{ab}(0) = 1.4 \div 2$ nm respectively [8], [14], [15]. These values give a depairing current density of about $j_d^{GL} = 1 \div 3 \times 10^8$ A/cm$^2$, a value close to the one extracted from the IVCs of nanowires capped with Au.

The difference between Jc values of nanowires with and without gold is a clear proof of the important role of the cap layer in determining the transport properties of our nanostructures.

A characteristic feature of the IVCs in Fig. 3(a) and 3(b) is the presence of a small step on the resistive branch, close to Ic (indicated by arrows in the two insets in Fig. 3). This feature could be explained in terms of phenomena like phase slip centers or hot spots, driving the superconducting channel to the normal state [16]. However, a systematic study to clarify its origin as a function of the wire width for samples with and without capping Au is in progress.

## IV. RESULTS AND DISCUSSIONS

In the previous section we have clearly shown that the protecting Au cap layer is a crucial factor for enhancing the quality of YBCO nanowires. To further prove this statement, we have analyzed and compared the resistance versus temperature measurements, R(T), for two nanowires with capping Au (Fig. 4) and without any capping (Fig. 5).

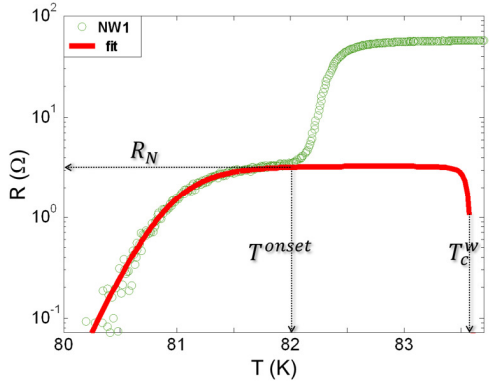

Fig. 4. Resistance versus temperature of the capped nanowire NW1 (width=65 nm, length=195 nm), along with a fit obtained using LAMH theory. Here we have fixed the wire dimensions and the value of the normal resistance $R_N$, extracting $T_c$, $\xi_0$ and $\lambda_0$.

In order to describe the broadening of the transition in these R(T) curves, we have supposed thermally activated phase slip events as the dominating source of resistance. We can thence study the transition using the Langer-Ambegaokar-McCumber-Halperin (LAMH) theory, describing the thermal activation of phase slips in nanowires [6], [7]. Although this theory has been developed for 1-dimentional systems ($w \leq \xi$) and is still affected by some contradictory results [17], [18], it has been demonstrated to give reasonable results also in the case of two-dimensional objects [19].

According to the LAMH theory, the finite resistance related to phase-slip processes $R_{LAMH}$ writes as

$$R_{LAMH}(T) = \left(\frac{h}{4e^2}\right)\left(\frac{\hbar\Omega_D}{k_B T}\right)\exp\left(-\frac{\Delta F}{k_B T}\right) \quad (2)$$

where $\Omega_D = (L/\xi)\sqrt{(\Delta F/k_B T)}\tau_{GL}^{-1}$ is the attempt frequency, $k_B$ is the Boltzmann constant, $L$ is the length of the wire, $\Delta F = (8\sqrt{2}/3)(B_c^2/2\mu_0)A\xi$ is the energy barrier for phase slip nucleation in a volume given by the cross sectional area of the wire $A$ times the coherence length $\xi$, $\tau_{GL} = \pi\hbar/8k_B(T_c^w - T)$ is the relaxation time. Close to $T_c^w$, the coherence length $\xi$ and critical field $B_c$ can be expressed $\xi(T) = \xi_0(1-T/T_c^w)^{-1/2}$ and $B_c(T) = B_c^0(1-T/T_c^w)$, where $B_c^0 = h/2e\sqrt{8}\pi\lambda_0\xi_0$ is the thermodynamical critical field, $\lambda_0$ and $\xi_0$ are the London penetration depth and the coherence length at T=0 K, respectively. The total resistance of the wire close to the transition can be expressed [5] as the parallel combination of $R_{LAMH}$ and the normal resistance of the wire $R_N$:

$$R(T) = (R_{LAMH}^{-1}(T) + R_N^{-1})^{-1} \quad (3)$$

In the following we will shortly discuss the effect of the normal resistance on the expected R(T) behavior and in detail the variation of the onset temperature of the resistive transition as a function of $R_N$.

In Figure 6(a) we show the calculated R(T) using equation (3) for 3 different values of $R_N$. Here we used fixed values for $\xi$, $\lambda$, $T_c$, L, and w, which are representative for our wire structures. The dashed line represents $R_N = \infty$. For a finite normal resistance $R_N=R_{N1}=80$ $\Omega$, which is approximately the normal resistance of our bare YBCO nanowires, we obtain the dash-dotted line. Decreasing the value of the normal resistance further, $R_N=R_{N2}<R_{N1}$ we obtain the solid line. One can clearly see the value of the onset temperature $T^{onset}$, at which the resistive transition will be observable, decreases with decreasing normal resistance value $R_N$. The value of $R_N$ in our Au capped nanowires can be approximated by the resistance of the gold strip of length $L_{Au} \simeq L_{wire}$, $R_{Au}^w$, on top of the YBCO wire.[1] This can be represented by the equivalent circuit shown in Fig 6(b) where the gold film is replaced by a resistive shunt $R_{Au}^w$ between the two wide electrodes.

We can see in fact from the R(T) measurements that the nanowires without Au (as the one in Fig. 5) show a rather sharp drop of resistance at about 85 K, close to the $T_c$ of the unpatterned YBCO film, corresponding to the transition of the electrodes. The transition of the wire is represented by a much broader foot, on whose onset $R_N$ and $T_c^w$ are defined. Instead for nanowires with gold capping (as the one in Fig. 4), because of the Au shunt, the onsets of the transition both of the electrodes and of the wires, not corresponding to the critical temperatures, are about 2K lower than the ones observed for the uncapped YBCO nanowires.

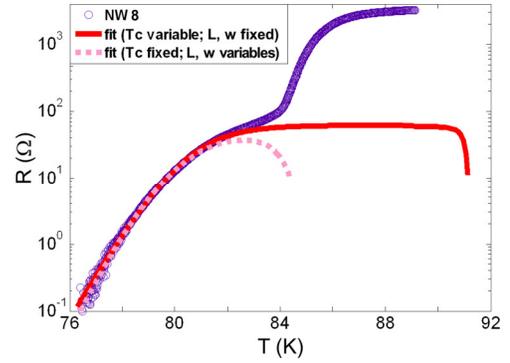

Fig. 5. Resistance versus temperature of the uncapped nanowire NW8 (width=43 nm, length=195 nm), along with the two fits using LAMH theory.

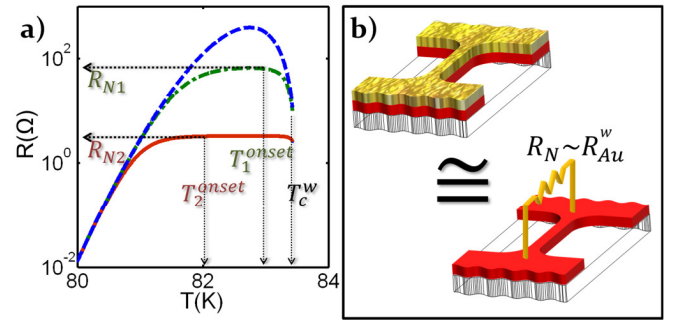

Fig. 6. (a) Influence of two normal resistances $R_{N1}$ and $R_{N2}$ on the resistance R of the nanowire described by the LAHM theory (dashed curve). The difference between the onset temperature and the critical temperature increases as the normal resistance is decreased. (b) The gold layer acts as a shunt for the YBCO nanowire: the normal resistance of the Au/YBCO wire can be thence approximated to the one of the gold strip on the top of the wire.

[1]We confirmed this approximation from simulations of the current distributions in our structures using typical values for the resistivity of Au and YBCO, and the contact resistivity between Au and YBCO.

TABLE I
FIXED AND EXTRACTED PARAMETERS FOR 9 NANOWIRES, 200 NM LONG (THE LAST 2 ARE UNCAPPED). THE $J_c$ VALUES HAVE BEEN MEASURED AT 4.2 K

| NW # | w (nm) | $R_N$ (Ω) | $T_c$ (K) | $\lambda_0$ (nm) | $\xi_0$ (nm) | $J_c$ (A/cm$^2$) GL theory | $J_c$ (A/cm$^2$) measured |
|---|---|---|---|---|---|---|---|
| 1 | 65  | 3.3  | 83.6 | 252 | 2.3 | 6.9e7 | 5.2e7 |
| 2 | 41  | 5.0  | 84.4 | 358 | 2.1 | 3.7e7 | 2.5e7 |
| 3 | 95  | 1.4  | 83.5 | 236 | 2.2 | 8.2e7 | 1.0e8 |
| 4 | 112 | 2.0  | 84.0 | 365 | 2.4 | 3.2e7 | 3.2e7 |
| 5 | 130 | 1.8  | 82.9 | 238 | 2.1 | 8.5e7 | 7.4e7 |
| 6 | 60  | 3.3  | 83.3 | 261 | 2.3 | 6.4e7 | 6.0e7 |
| 7 | 120 | 2.3  | 82.9 | 207 | 2.1 | 1.1e8 | 8.3e7 |
| 8 | 43  | 63.3 | 91.4 | 554 | 2.4 | 1.4e7 | 2.0e7 |
| 9 | 116 | 51.0 | 88.1 | 771 | 1.8 | 9.4e6 | 4.2e6 |

The R(T) curves of the capped nanowires have been fitted according to eq. (3), as shown in Fig. 4. For the fitting procedure we have fixed the normal resistance $R_N$ as the resistance value at the onset temperature $T^{onset}$ and the cross section and the length of the wire measured by AFM, and we have considered $T_c^w$, $\xi_0$ and $\lambda_0$ as fitting parameters. We got a $T_c^w$ value of 83.6 K, which is quite consistent with the $T_c$ of the uncapped wires and that of the unpatterned YBCO films; for the coherence length and London penetration depth we obtained values of 2.3 nm and 252 nm respectively, in agreement with typical values for YBCO [8], [20].

As a further confirmation of the validity of the fitting procedure to extract physical values of $\xi$ and $\lambda$ representing the nanowire, we have calculated the expected current density of the nanowire with eq (1) using the values of $\xi_0$ and $\lambda_0$ obtained by the fit. The expected $J_c$ so calculated is consistent with the value extracted from the IV curve. This result, obtained here for the first time for YBCO nanowires, proves once more the high quality of our nanostructures, preserving "pristine" superconducting properties because of the presence of the gold capping acting as a resistive shunt. Here it's worth mentioning that, despite the presence of an external shunt resistance has been recently associated to the occurrence of a critical current enhancement [21], our gold mainly acts like a protective capping. The R(T) curves of 6 more nanowires capped with gold have been successfully fitted using the same procedure. A summary of the results is reported in table I.

The scenario is completely different for uncapped nanowires. We have tried to fit the R(T) curves using equation (3), by either fixing $T_c^w$ extracted from the measurement or considering it a fitting parameter. In Fig. 5 (solid line) the R(T) of the nanowire NW8, 43nm wide and 200 nm long, has been fitted - as previously done for capped wires - considering $T_c^w$ as a fitting parameter: the foot structure can be well fitted, and the extracted $\xi_0$ and $\lambda_0$ (2.4 nm and 554 nm respectively) are consistent with the measured critical current density. However, we obtained a completely unfeasible $T_c^w$ of more than 91 K, much higher than the critical temperature of the bare YBCO film.

The same R(T) has been fitted (Fig. 5, dashed line) by imposing $T_c^w$ close to 84K, a temperature where the onset of the wire transition is visible. In this case, in order to fit the data we need to consider the cross sectional area A and the length L as free parameters and to fix the values of $\xi_0$ and $\lambda_0$ close to the ones previously extracted for gold capped wires (2 nm and 200 nm, respectively). We obtained A=353 nm$^2$ and L=6.1 nm, respectively 6 and 33 times smaller than the nominal cross section and length, measured by AFM. This discrepancy is certainly related to the local damage of the wire during fabrication. As a consequence, the LAMH theory can be hardly applied to uncapped wires.

## V. CONCLUSION

We have implemented a careful nanopattening procedure, fabricating 50 nm thick YBCO nanowires, capped with gold, having widths and lengths in the range 40-200 nm and 200-3000 nm, respectively. The effective role of the cap layer is confirmed by current voltage characteristics (IVCs) and resistance versus temperature R(T) measurements. The nanowires carry in fact critical current densities up to $10^8$ A/cm$^2$, approaching the theoretical depairing current limit. Moreover, the resistance close to the superconducting transition can be studied in terms of the LAMH theory describing the thermal activation of phase slips in nanowires and the extracted $\xi_0$ and $\lambda_0$ give a value of Jc consistent with the measured critical current density. If the gold cap layer is removed, the nanowires are affected by local damages and their R(T) can't be explained in terms of TAPS model unless considering wire dimensions far from those measured by AFM. Our nanowires can be a good candidate for the study of superconducting phenomena in HTS and for their integration in magnetometers and devices for photodetection experiments.